\documentclass{eas}
\usepackage{graphicx}
\usepackage{natbib}
\TitreGlobal{Ecole Evry Schatzman 2010: Star Formation in the Local
Universe}
\begin{document}

\title{Star Formation on Galactic Scales: Empirical Laws}
\author{Bruce G. Elmegreen}\address{IBM T. J. Watson Research Center, 1101 Kitchawan Road, Yorktown
Heights, New York 10598 USA, bge@us.ibm.com}
\begin{abstract}
Empirical star formation laws from the last 20 years are reviewed with
a comparison to simulations. The current form in main galaxy disks has
a linear relationship between the star formation rate per unit area and
the molecular cloud mass per unit area with a timescale for molecular
gas conversion of about 2 Gyr. The local ratio of molecular mass to
atomic mass scales nearly linearly with pressure, as determined from
the weight of the gas layer in the galaxy. In the outer parts of
galaxies and in dwarf irregular galaxies, the disk can be dominated by
atomic hydrogen and the star formation rate per unit area becomes
directly proportional to the total gas mass per unit area, with a
consumption time of about 100 Gyr. The importance of a threshold for
gravitational instabilities is not clear. Observations suggest such a
threshold is not always important, while simulations generally show
that it is. The threshold is difficult to evaluate because it is
sensitive to magnetic and viscous forces, the presence of spiral waves
and other local effects, and the equation of state.
\end{abstract}
\maketitle

\section{Introduction: the Kennicutt-Schmidt law of Star Formation}

Star formation and stellar evolution are such important drivers of
galactic evolution that empirical laws to determine the star formation
rate have been investigated for over 50 years. The results have never
been very precise because star formation spans a wide range of scales,
from cluster-forming cores to molecular clouds to the whole
interstellar medium.

On the scale of a galaxy, the first idea was a proposed connection
between the total star formation rate and the mass of interstellar gas.
\cite{schmidt59} derived the star formation rate (SFR) over the history
of the Milky Way assuming a constant initial luminosity function for
stars, $\Psi(M_{\rm V})$, a stellar lifetime function $T(M_{\rm V})$, a
gas return per star equal to all of the stellar mass above
$0.7\;M_\odot$, and a star formation rate $f(t)$ that scales with a
power $n$ of the gas mass, $M_{\rm G}(t)$. Then $f(t)\Sigma_{\rm MV}
\Psi(M_{\rm V}) = C [M_{\rm G}(t)]^n$, for a summation $\Sigma_{\rm
MV}$ over all stellar types.

Schmidt gave analytical solutions for $n=0,1,2$. He noted that a scale
height for {\rm HI} of 144 pc, a scale height for Cepheids of 80 pc,
and a scale height for clusters of 58 pc gave $n=2$ to 3.  The white
dwarf count gave $n>2$, the He abundance suggested $n=2$, the
uniformity of {\rm HI} suggested $n\ge2$, and the cluster mass function
gave $n=1$ to 2. Schmidt also suggested that with $n=2$, dense galaxies
like ellipticals should now have less gas than low-density galaxies
like the LMC. His final comment was ``It is hoped to study the
evolution of galaxies in more detail in the future.''  Following
\cite{schmidt59}, many authors derived scaling relations between the
average surface density of star formation, $\Sigma_{\rm SFR}$, and the
average surface density of gas. \cite{buat89} included molecular and
atomic gas and determined star formation rates from the UV flux
corrected for Milky Way and internal extinction. They assumed a
constant ${\rm H}_{\rm 2}/{\rm CO}$ ratio and a \cite{scalo86} IMF. The
result was a good correlation between the average star formation rate
in a sample of 28 galaxies and the $1.65\pm0.16$ power of the average
total gas surface density. In the same year, \cite{ken89} used
H$\alpha$ for star formation, and {\rm HI} and {\rm CO} for the gas
with a constant ${\rm H}_{\rm 2}/{\rm CO}$ conversion factor, and
determined star formation rates both as a function of galactocentric
radius and averaged over whole galaxy disks. For whole galaxies, the
average H$\alpha$ flux scaled with the average gas surface density to a
power between 1 and 2; there was a lot of scatter in this relation and
the correlation was better for {\rm HI} than ${\rm H}_{\rm 2}$. More
interesting was Kennicutt's (1989) result that the star formation rate
had an abrupt cutoff in radius where the \cite{toomre64} stability
condition indicated the onset of gravitationally stable gas. Kennicutt
derived a threshold gas column density for star formation, $\Sigma_{\rm
crit}=\alpha\sigma\kappa/(3.36G)$ for $\alpha=0.7$; $\sigma$ is the
velocity dispersion of the gas; $\kappa$ is the epicyclic frequency,
and $G$ is the gravitational constant.

In a second study, \cite{ken98} examined the disk-average star
formation rates using a larger sample of galaxies with H$\alpha$, {\rm
HI}, and {\rm CO}. He found that for normal galaxies, the slope of the
SFR-surface density relation ranged between 1.3 to 2.5, depending on
how the slope was measured; there was a lot of scatter. When starburst
galaxies with molecular surface densities in excess of 100 $M_\odot$
were included, the overall slope became better defined and was around
1.4.  This paper also found a good correlation with a star formation
rate that scaled directly with the average surface density of gas and
inversely with the rotation period of the disk. This second law
suggested that large-scale dynamical processes are involved.

\cite{hunter98} considered the same type of analysis for dwarf
Irregulars and derived a critical surface density that was lower than
the \cite{ken89} value by a factor of $\sim2$. This meant that stars
form in more stable gas in dwarf irregulars compared to spirals.

\cite{boissier03} compared $\Sigma_{\rm SFR}$ and $\Sigma_{\rm gas}$
versus radius in 16 resolved galaxies with three theoretical
expressions. The best fits were a SFR dependence on the gas surface
density as $\Sigma_{\rm SFR}\propto\Sigma_{\rm gas}^{2.06}$, a more
dynamical law from \cite{bp99} which gave the fit $\Sigma_{\rm
SFR}\propto\Sigma_{\rm gas}^{1.48}(V/R)$ for rotation speed $V$ and
radius $R$, and a third type of law from \cite{dopita94}, which fit to
$\Sigma_{\rm SFR}\propto\Sigma_{\rm gas}^{0.97}/\Sigma_{\rm
tot}^{0.61}$. \cite{boissier03} assumed that H$_{\rm 2}$/{\rm CO}
varied with radius as the metallicity \citep{boselli02}. Their
conclusion was that the three laws are equally good, and that for the
pure gas law, $n>1.4$. \cite{boissier03} also looked for a star
formation threshold in the Milky Way. They determined
$\Sigma/\Sigma_{\rm crit}$ using both pure-gas for $\Sigma_{\rm crit}$
and a gas+star $\Sigma_{\rm crit}$ from \cite{ws94}. They found that
the gas+star $\Sigma_{\rm crit}$ gave the best threshold for
determining where star formation occurs. The gas alone was
sub-threshold throughout the disk.

\cite{zs05} showed that a threshold like $\Sigma_{\rm crit}$ may be
used to determine the gas fraction in galaxies. If all galaxies have
$\Sigma({\rm HI})$ approximately at the critical $\Sigma_{\rm crit}
=\alpha \kappa \sigma/\pi G$, which is proportional to $V/R$ from
$\kappa$, then $M_{\rm gas} = \int_R 2\pi R\Sigma_{\rm crit}dR \propto
VR$. This was shown to be the case from observations. They also
considered that the total mass is $M_{\rm tot}\propto V^2R$, in which
case $M_{\rm tot}/M_{\rm gas}\propto V$, the rotation speed. This was
also shown to be confirmed by observations. In their interpretation,
small galaxies are more gas-rich than large galaxies because all
galaxies have their gas column densities close to the surface density
threshold.

For the Milky Way, \cite{mis06} used {\rm CO}BE/DIRBE observations to
get both the gas and dust distributions and the SFR distribution. They
found a gas-law slope of $2.18\pm0.20$, which they claimed was similar
to Kennicutt's (1998) bivariate fit slope $n=2.5$ for normal galaxies.
\cite{luna06} determined the Milky Way SFR from IRAS point sources and
the {\rm CO} surface density from a southern hemisphere survey
(assuming constant H$_{\rm 2}$/{\rm CO}). They found star formation
concentrated in low-shear spiral arms and suggested an additional
dependence on shear. Overall they derived $\Sigma_{\rm
SFR}\sim\Sigma_{\rm gas}^{1.2\pm0.2}$. \cite{voro03} also suggested a
shear dependence for the SFR based on observations of the Cartwheel
galaxy, where there is an inner ring of star formation with high shear
that is too faint for the normal Kennicutt law, given the gas column
density.

\subsection{The Q Threshold}\label{sect:q}

A threshold for gravitational instabilities in rotating disks has been
derived for various ideal cases. For an infinitely thin disk of
isothermal gas, the dispersion relation for radial waves is $\omega^2 =
k^2\sigma^2 - 2\pi G\Sigma k + \kappa^2$. Solving for the fastest
growth rate $\omega$ gives the wavenumber at peak growth, $k = \pi
G\Sigma/\sigma^2$, and the wavelength, $\lambda=2\sigma^2/G\Sigma$,
which is on the order of a kiloparsec in main galaxy disks. The
dominant unstable mass is
$M\sim(\lambda/2)^2\Sigma=\sigma^4/G^2\Sigma\sim10^7\;M_\odot$ in local
spirals. The peak rate is given by
\begin{equation} \omega_{peak}^2 = -(\pi G\Sigma/\sigma^2)^2 + \kappa^2 =
-(\pi G \Sigma/\sigma^2)^2(1-Q^2)\end{equation} which requires
$Q\equiv\kappa\sigma/\pi G\Sigma < 1$ for instability (i.e., when
$\omega_{peak}^2<0$).

Disk thickness weakens the gravitational force in the in-plane
direction by an amount that depends on wavenumber, approximately as
$1/(1+kH)$ for exponential scale height $H$ \citep[e.g.,][]{e87, ko07}.
Typically, $k\sim1/H$, so this weakening can slow the instability by a
factor of $\sim2$, and it can make the disk slightly more stable by a
factor of 2 in $Q$. On the other hand, cooling during condensation
decreases the effective value of the velocity dispersion, which should
really be written $\gamma^{1/2}\sigma$ for adiabatic index $\gamma$
that appears in the relation $\delta P\propto\delta \rho^\gamma$ with
pressure $P$ and density $\rho$.  If $P$ is nearly constant for changes
in $\rho$, as often observed, then $\gamma\sim0$. \cite{myers78} found
$\gamma\sim0.25$ for various thermal temperatures at interstellar
densities between 0.1 cm$^{-3}$ and 100 cm$^{-3}$.  Thus the effects of
disk thickness and a soft equation of state partially compensate for
each other.

There is also a $Q$ threshold for the collapse of an expanding shell of
gas \citep{epe02}. Pressures from OB associations form giant shells of
gas and cause them to expand. Eventually they go unstable when the
accumulated gas is cold and massive enough, provided the induced
rotation and shear from Coriolis forces are small. Considering
thousands of initial conditions, these authors found that a sensitive
indicator of whether collapse occurs before the shell disperses is the
value of $Q$ in the local galaxy disk, i.e., independent of the shell
itself. The fraction $f$ of shells that collapsed scaled inversely with
$Q$ as $f\sim0.5- 0.4\log_{10} Q$.

The Toomre $Q$ parameter is also likely to play a role in the
occurrence of instabilities in turbulence-compressed gas on a galactic
scale \citep{e02}. Isothermal compression has to include a mass
comparable to the ambient Jeans mass, $M_{\rm Jeans}$, in order to
trigger instabilities. The turbulent outer scale in the galaxy is
comparable to the Jeans length, $L_{\rm Jeans}$, which is about the
galactic gas scale height, $H$. If the compression distance exceeds the
epicyclic length, then Coriolis forces spin up the compressed gas,
leading to resistance from centrifugal forces. So instability needs
$L_{\rm Jeans}\le L_{\rm epicycle}$, which means $Q\le1$, since $L_{\rm
Jeans}\sim H \sim \sigma^2/\pi G\Sigma$. The epicyclic length is
$L_{\rm epicycle}\sim\sigma/k$, so $L_{\rm Jeans}/L_{\rm epicycle}=Q$.

The dimensionless parameter $Q$ measures the ratio of the centrifugal
force from the Coriolis spin-up of a condensing gas perturbation to the
self-gravitational force, on the scale where gravity and pressure
forces are equal, which is the Jeans length. The derivation of $Q$
assumes that angular momentum is conserved, so the Coriolis force spins
up the gas to the maximum possible extent. When $Q>1$, a condensing
perturbation on the scale of the Jeans length spins up so fast that its
centrifugal force pulls it apart against self-gravity. Larger-scale
perturbations have the same self-gravitational acceleration (which
scales with $\Sigma$) and stronger Coriolis acceleration (which scales
with $\kappa^2/k$); smaller-scale perturbations have stronger
accelerations from pressure. If angular momentum is not conserved, then
the disk can be unstable for a wider range of $Q$ because there is less
spin up during condensation.  For example, the Coriolis force can be
resisted by magnetic tension or viscosity and then the angular momentum
in a condensing cloud will get stripped away. This removes the $Q$
threshold completely \citep{chandra54,stephenson61,lyndenbell66,hh83}.
In the magnetic case, the result is the Magneto-Jeans instability,
which can dominate the gas condensation in low-shear environments like
spiral arms and some inner disks \citep{e87, e91, e94, ko01, ko02,
kos02}. For the viscous case, \cite{gammie96} showed that for $Q$ close
to but larger than 1, i.e., in the otherwise stable regime, viscosity
can make the gas unstable with a growth rate equal to nearly one-third
of the full rate for a normally unstable ($Q<1$) disk. A dimensionless
parameter for viscosity $\nu$ is $\nu\kappa^3/G^2\Sigma^2$, which is
$\sim11$ according to \cite{gammie96}. This is a large value indicating
that galaxy gas disks should be destabilized by viscosity. An important
dimensionless parameter for magnetic tension is $B^2/(\pi
G\Sigma^2)\sim8$, which is also large enough to be important. Thus gas
disks should be generally unstable to form small spiral arms and
clouds, even with moderately stable $Q$, although the growth rate can
be low if $Q$ is large.

\subsection{Modern Versions of the KS Law with $\sim1.5$ slope}

\cite{ken07} studied the local star formation law in M51 with 0.5-2 kpc
resolution using Pa-$\alpha$ and 24$\mu$+H$\alpha$ lines for the SFR,
and a constant conversion factor for {\rm CO} to H$_{\rm 2}$. There was
a correlation, mostly from the radial variation of both SFR and gas
surface density, with a slope of $1.56\pm0.04$. There was no
correlation with $\Sigma({\rm HI})$ alone, as this atomic component had
about constant column density ($\sim10\;M_\odot$). The correlation with
molecules alone was about the same as the total gas correlation.

\cite{leroy05} studied dwarf galaxies and found that they have a
molecular KS index of $1.3\pm0.1$, indistinguishable from that of
spirals, except with a continuation to lower central ${\rm H}_{\rm 2}$
column densities (i.e., down to $\sim10\;M_\odot$ pc$^{-2}$).

\cite{heyer04} found a slope $n=1.36$ for $\Sigma_{\rm SFR}$ versus
$\Sigma({\rm H}_{\rm 2})$ in M33, where the molecular fraction, $f_{\rm
mol}$ is small. The correlation with the total gas was much steeper.
More recently, \cite{verley10} studied M33 again and got $\Sigma_{\rm
SFR}\propto\Sigma_{\rm H2}^n$ for $n=1$ to 2, and $\Sigma_{\rm
SFR}\propto\Sigma_{\rm total\;gas}^n$ for $n=2$ to 4. The steepening
for total gas is again because $\Sigma_{\rm {\rm HI}}$ is about
constant, so the slope from {\rm HI} alone is nearly infinite. This
correlation is dominated by the radial variations in both quantities,
as it is a point-by-point evaluation throughout the disk. Radial
changes in metallicity, spiral arm activation, tidal density, and so
on, are part of the total correlation.  \cite{verley10} also try other
laws, such as $\Sigma_{\rm SFR}\propto \left(\Sigma_{\rm H2} \rho_{\rm
ISM}^{0.5}\right)^n$, for which $n=1.16\pm0.04$, and $\Sigma_{\rm
SFR}\propto\rho_{\rm ISM}^n$, for which $n=1.07\pm0.02$. These differ
by considering the conversion from column density to midplane density,
using a derivation of the gaseous scale height. The first of these
would have a slope of unity if the star formation rate per unit
molecular gas mass were proportional to the dynamical rate at the
average local (total) gas density. The second has the form of the
original Schmidt law, which depends only on density. To remove possible
effects of {\rm CO} to H$_{\rm 2}$ conversion, Verley et al. also
looked for a spatial correlation with the 160 $\mu$ opacity,
$\tau_{160}$, which is a measure of the total gas column density
independent of molecule formation. They found $\Sigma_{\rm
SFR}\propto\tau_{160}^n$ for $n=1.13\pm0.02$, although the correlation
was not a single power law but a 2-component power law with a shallow
part (slope $\sim0.5$) at low opacity ($\tau_{160}<10^{-4}$) and a
steep part (slope $\sim2$) at high opacity.

\subsection{Explanations for the 1.5 slope}

Prior to around 2008, the popular form of the KS law had a slope of
around 1.5 when $\Sigma_{\rm SFR}$ was plotted versus total gas column
density on a log-log scale. This follows from a dynamical model of star
formation in which the SFR per unit area equals the available gas mass
per unit area multiplied by the rate at which this gas mass gets
converted into stars, taken to be the dynamical rate,
\begin{equation}
\Sigma_{\rm SFR}\sim \epsilon\Sigma_{\rm gas} \left(G\rho_{\rm
gas}\right)^{1/2}.
\end{equation}
If the gas scale height is constant, then $\Sigma_{\rm
gas}\propto\rho_{\rm gas}$ and $\Sigma_{\rm SFR}\propto\Sigma_{\rm
gas}^{1.5}$. In the model of star formation where star-forming clouds
are made by large-scale gravitational instabilities, this 1.5 power law
would work only where the Toomre instability condition, $Q\leq1.4$, is
satisfied. Such a model accounts for the Kennicutt (1989, 1998) law
with the $Q<1.4$ threshold.

Several computer simulations have shown this dynamical effect.
\cite{li06} did SPH simulations of galaxy disks with self-gravity
forming sink particles at densities larger than $10^3$ cm$^{-3}$.  They
found a $Q$ threshold for sink particle formation, and had a nice fit
to the KS law with a slope of $\sim1.5$. \cite{tb06} ran ENZO, a 3D
adaptive mesh code, with star formation at various efficiencies,
various temperature floors in the cooling function, and various
threshold densities. Some models had a low efficiency with a low
threshold density and other models had a high efficiency with a high
threshold density. Some of their models had feedback from young stars.
They also got a KS slope of $\sim1.5$ for both global and local star
formation, regardless of the details in the models. \cite{kravtsov03}
did cosmological simulations using N-body techniques in an Eulerian
adaptive mesh. He assumed a constant efficiency of star formation at
high gas density, and star formation only in the densest regions
($n>50$ cm$^{-3}$, the resolution limit), which are in the tail of the
density probability distribution function \citep[pdf; cf.][]{e02,
km05}. \cite{kravtsov03} got the KS law with a slope of 1.4 for total
gas surface density. \cite{wn07} did a similar thing, using the
fraction of the mass at a density greater than a critical value from
the pdf ($\rho_{\rm crit}=10^3$ cm$^{-3}$) to determine the star
formation rate. Their analytical result had a slope of 1.5.
\cite{harfst06} had a code with a hierarchical tree for tracking
interacting star particles, SPH for the diffuse gas, and sticky
particles for the clouds. They included mass exchange by condensation
and evaporation, mass exchange from stars to clouds (via PNe) and from
stars to diffuse gas (SNe), and from clouds into stars during star
formation. New clouds were formed in expanding shells. Their KS slope
was $1.7\pm0.1$. They also got a drop in $\Sigma_{\rm SFR}$ at low
$\Sigma_{\rm gas}$, not from a $Q$ threshold but from an inability of
the gas to cool and form a thin disk \citep[cf.][]{burkert92,ep94}.

\section{The Molecular Star Formation Law}

\begin{figure}[b]
\begin{center}
 \includegraphics[width=5.in]{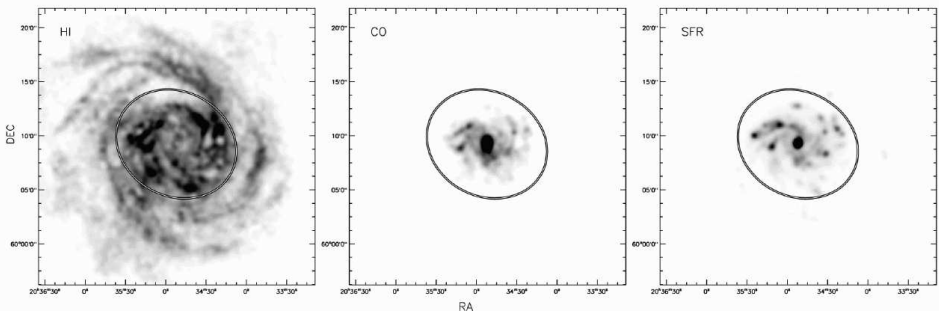}
\caption{Maps of {\rm HI}, {\rm CO} and SFR in NGC 6946 with {\rm HI}
on the left, {\rm CO} in the middle and SFR on the right, all convolved
to 750 pc resolution (from \cite{bigiel08}. The circle is the optical
radius at 25 mag arcsec$^{-2}$.}\label{f1}
\end{center}
\end{figure}

The star formation law may also be written as a linear relation for
molecules, with $\Sigma_{\rm SFR}\propto\Sigma_{\rm H2}^{1}$
\citep[e.g.,][]{ry99}. \cite{wb02} found a SFR in direct proportion to
molecular cloud density ($n=1$), and suggested that the $n=1.4$ KS law
came from changes in the molecular fraction, $f_{\rm mol}=\Sigma_{\rm
H2}/\left(\Sigma_{\rm {\rm HI}}+\Sigma_{\rm H2}\right)$. They assumed
that H$_{\rm 2}$/{\rm CO} was constant and determined the combined
index $n^\prime=n_{\rm mol}\left(1+d\ln f_{\rm mol}/d\ln\Sigma_{\rm
gas}\right)$ where $n_{\rm mol}=1$ and $f_{\rm mol}$ increases with
pressure, $P$. They measured $d\ln f_{\rm mol}/d\ln P \sim 0.2$, and if
$P \propto \Sigma_{\rm gas}^2$, then $d\ln f_{\rm mol}/d\ln\Sigma_{\rm
gas}\sim 0.4$. This gives the KS $n=1.4$ law for total gas.  Wong \&
Blitz also suggested that the stability parameter $Q$ was not a good
threshold for star formation, but a better measure of the gas fraction
in the sense that a high $Q$ corresponds to a low $\Sigma_{\rm
gas}/\Sigma_{\rm tot}$. \cite{br06} showed for a wider sample of 13
galaxies that the molecular ratio, $R_{\rm mol}= \Sigma_{\rm
H2}/\Sigma_{\rm {\rm HI}}$, scales about linearly with the total ISM
pressure. Interacting galaxies had slightly higher $R_{\rm mol}$ for a
given $P$, but among interacting galaxies, the correlation was still
present.

\begin{figure}[b]
\begin{center}
 \includegraphics[width=3.in]{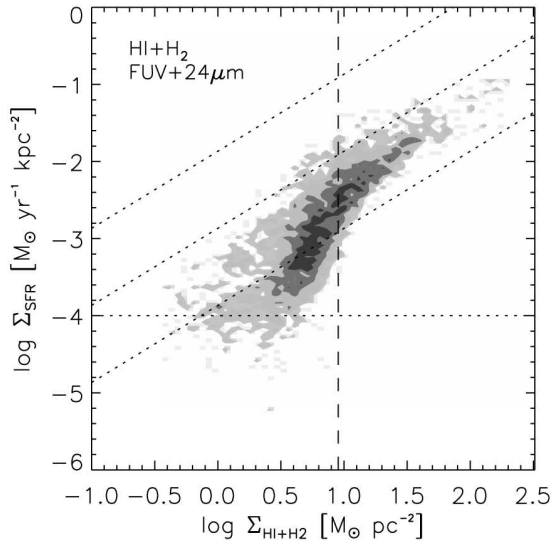}
\caption{The distribution of SFR per unit area versus total gas column
density, convolved to 750 pc, for 7 spiral galaxies (from
\cite{bigiel08}. There is a change in the slope from $\sim4$ at
$\Sigma_{{\rm HI+H2}}<9\;M_\odot$ pc$^{-2}$ (the vertical dashed line)
in the outer disk to $\sim1$ at higher $\Sigma_{{\rm HI+H2}}$ in the
inner disk. The short-dashed lines correspond to gas depletion times of
0.1 Gyr, 1 Gyr, and 10 Gyr, from top to bottom.  }\label{f2}
\end{center}
\end{figure}

A large study of {\rm HI}, {\rm CO}, and star formation rates from
GALEX ultraviolet and Spitzer 24$\mu$ observations was made by
\cite{bigiel08} and \cite{leroy08}. They considered the local star
formation law with a resolution of 750 pc. Bigiel et al. found that
$\Sigma_{\rm SFR}\propto\Sigma_{\rm CO}$, and that the timescale for
conversion from ${\rm H}_{\rm 2}$ to stars was about 2 Gyr. Figure
\ref{f1} \cite[from][]{bigiel08} shows an example of how much better
the SFR scales with {\rm CO} than either {\rm HI} or the total gas. The
{\rm CO} and SFR maps of NGC 6946 resemble each other closely, and
neither resembles the {\rm HI} map. Bigiel et al. also found that
$\Sigma_{\rm HI}$ saturates to $\sim9\;M_\odot$ pc$^{-2}$. When
plotting $\Sigma_{\rm SFR}$ over a wide range of $\Sigma_{{\rm
HI+H2}}$, they found a slope of unity in the molecular range,
$\Sigma_{{\rm HI+H2}}>9\;M_\odot$ pc$^{-2}$, and higher slope in the
atomic range ($\Sigma_{{\rm HI+H2}}<9\;M_\odot$ pc$^{-2}$). Figure
\ref{f2} shows the summed distribution of SFR per unit area versus
total gas column density in 7 spiral galaxies. There is a linear part
at high column density and a steeper part at low column density.

Dwarf galaxies look like the outer parts of spirals in the Bigiel et
al. survey, occupying the steeper part of the $\Sigma_{\rm
SFR}-\Sigma_{\rm gas}$ diagram at low $\Sigma_{\rm gas}$. At higher
$\Sigma_{{\rm HI+H2}}$, the survey did not have new data, but Bigiel et
al. suggested, based on Kennicutt's (1998) starburst result, that
perhaps the KS law turned up to a steeper slope ($n\sim1.4$) in a third
regime of star formation where $\Sigma_{\rm H2}$ exceeds the standard
column density of a single molecular cloud (around $100\;M_\odot$
pc$^{-2}$).

\cite{leroy08} compared these new survey results to various theoretical
models. They found that the star formation time in {\rm CO}-rich gas is
universally 1.9 Gyr, independent of the average local free fall or
orbital time, the midplane gas pressure, the state of gravitational
stability of the disk with or without the inclusion of stars in the
stability condition, and regardless of the rate of shear or the ability
of a cold gas phase to form.  Star formation depends only on the
presence of molecules and it proceeds at a fixed rate per molecule.
Leroy et al. also found that dwarf galaxies are forming stars at their
average historical rate, whereas spirals are forming stars at about
half of their average rate. In the outer disk, the SFR in {\rm HI}
drops with radius faster than the free fall time, suggesting
self-gravity is not the lone driver. Also important are the phase
balance between {\rm HI} and H$_{\rm 2}$, giant molecular cloud (GMC)
destruction, stellar feedback, and other processes. These processes
govern the presence of GMCs with an apparently constant star formation
efficiency in each GMC.

Unlike the star formation rate per molecule, the molecule-to-atom ratio
does correlate well with environmental parameters. \cite{leroy08}
showed approximately linear correlations with stellar surface density
and interstellar pressure, an inverse squared dependence on the orbit
time, and an exponential dependence on the galactic radius, like the
rest of the disk, and with a comparable radial scale length.  The
molecular fraction is a smooth function of environmental parameters
(e.g., pressure); no thresholds were seen. Disks seem to be marginally
stable throughout.

Leroy et al. concluded by noting that the {\rm HI}-H$_{\rm 2}$
transition in spirals typically occurs at $0.43\pm0.18\;R_{\rm 25}$,
which is about the same as where $\Sigma_{\rm stars}=81\pm25\;
M_\odot\; {\rm pc}^{-2}$, $\Sigma_{\rm gas}=14\pm6\; M_\odot\;{\rm
pc}^{-2}$, $P=2.3\pm1.5\times10^4k_{\rm B}\; {\rm K}\; {\rm cm}^{-3}$,
and $T_{\rm orbit}=1.8\pm0.4$ Gyr. There should be unobserved ${\rm
H}_{\rm 2}$ in dwarfs, according to the high star formation rates and
low {\rm CO} emissions there; in fact Leroy et al. estimate for dwarfs
$\Sigma_{\rm H2}\sim2\Sigma_{\rm {\rm HI}}$ in the inner regions.

Where $\Sigma_{\rm {\rm HI}}>\Sigma_{\rm H2}$, the star formation
efficiency is proportional to $\Sigma_{\rm stars}$, making
\begin{equation}
\Sigma_{\rm SFR}\sim \Sigma_{\rm gas}\times \left[\Sigma_{\rm
stars}/81\;M_\odot\;{\rm pc}^{-2}\right]/1.9\;{\rm Gyr}.
\end{equation}

\subsection{Theoretical Models for the Bigiel-Leroy Observations}

\cite{kmt08} considered the molecule formation problem by starting with
the radiative transfer of ${\rm H}_{\rm 2}$-dissociating radiation:
$dF/dz = -n\sigma_{\rm d} F - f_{\rm {\rm HI}}n^2R/f_{\rm diss}$. Here,
$F$ is the flux in Lyman-Werner bands that dissociate H$_{\rm 2}$, $n$
is the density ($\sim30$ cm$^{-3}$ near the H$_{\rm 2}$ transition),
$\sigma_{\rm d}$ is the dust cross section per H ($10^{-21}$ cm$^2$),
$f_{\rm {\rm HI}}$ is the fraction of $n$ that is {\rm HI}, $R$ is the
rate coefficient for formation of H$_{\rm 2}$ on grains
($\sim3\times10^{-17}$ cm$^3$ s$^{-1}$; i.e., the formation rate is
$f_{\rm {\rm HI}}n^2R$), and $f_{\rm diss}$ is the fraction of uv
photon absorptions that dissociate H$_{\rm 2}$ ($\sim0.1$).

The solution to this radiative transfer equation is $F(\tau) =
\left(e^{-\left[\tau-\tau_{\rm {\rm HI}}\right]} -1\right)/\chi$ where
$\tau_{\rm {\rm HI}}=\ln(1+\chi)$; $\chi = f_{\rm diss} \sigma_{\rm d}
c E_{\rm 0}/nR$ is the ratio of absorption in dust to H$_{\rm 2}$, and
$E_{\rm 0}$ is the free space photon number density
($\sim7.5\times10^{-4}$ cm$^{-3}$). Krumholz et al. (2008) assume that
the cold neutral density comes from two phase equilibrium. Then $n$
scales with $E_{\rm 0}$ and $\chi$ becomes nearly constant.  From this
they get the extinction, $A_{\rm V}$, to the {\rm HI}/H$_{\rm 2}$
transition, the {\rm HI} column density, $\Sigma_{\rm {\rm HI}}$, and
the molecular fraction in spherical cloud complexes as a function of
the complex total column density. They do this also as a function of
metallicity.

After considering a galactic cloud population, \cite{kmt09a} derive
$\Sigma_{\rm {\rm HI}}$ versus $\Sigma_{\rm total\;gas}$ for different
metallicities, and compare this with observations of galaxies having
those metallicities. They do the same for H$_{\rm 2}$. They also
compare the observed versus the predicted correlation between H$_{\rm
2}$/{\rm HI} and pressure $P$. To do this, they use the observed
$\Sigma_{\rm total\;gas}$ and metallicity, and then compute $R_{\rm
H2}=\Sigma_{{\rm H}_{\rm 2}}/\Sigma_{\rm {\rm HI}}$ from theory. This
is plotted versus the observed pressure from \cite{br06} and
\cite{leroy08}.  The agreement is good.

\cite{kmt09b} considered the star formation law,
\begin{equation} \Sigma_{\rm SFR} = \Sigma_{\rm gas}f_{\rm H2} SFR_{\rm ff}/t_{\rm ff}
\end{equation}
where  the star formation rate in a free fall time is the fraction of
the gas that turns into stars in a free fall time, $SFR_{\rm ff}$,
divided by the free fall time, $t_{\rm ff}$. This is $SFR_{\rm
ff}/\tau_{\rm ff} = (M_6^{-0.33}/0.8\;{\rm Gyr})\times {\rm
Max}[1,\Sigma_{\rm gas} / 85\;M_\odot\;{\rm pc}^{-2}]$; $M_6$ is the
cloud mass in units of $10^6\;M_\odot$. This equation assumes that
stars form in the high density tail of a log-normal density pdf, with
the tail width given by the Mach number; a fraction of 0.3 of the dense
gas mass goes into stars. The clouds are virialized and at uniform
pressure until the galactic $\Sigma_{\rm gas}$ exceeds the column
density of a single GMC; then the pressure equals the galactic
pressure. Also, the cloud complex mass is taken to be
$M_6=37\Sigma_{\rm gas}/(85\;M_\odot\;{\rm pc}^{-2})$ from the galaxy
Jeans mass.

This theory for molecule formation and star formation in a galactic
environment fits well to the observations by \cite{bigiel08} and
\cite{leroy08}. It reproduces the low column density regime by having a
low ratio of molecules to atoms at low pressure, it reproduces the
intermediate column density regime by having a fixed star formation
rate per molecule and an areal average star formation rate from the
areal density of molecular clouds at constant pressure, and it
reproduces the high column density regime by increasing the
interstellar pressure, which makes the cloud density go up and the free
fall time go down. A key point in their model is that molecular cloud
pressures are constant in normal galaxy disks because they are set by
{\rm HI}I region pressures (feedback) and not the galactic environment.
In this sense, all GMCs have to be parts of shells or other active
disturbances formed by high pressures.

We know that molecular cloud pressures in the Milky Way are about
constant from the \cite{larson81} laws, which require this for
virialized clouds, but we don't really know the reason for it. It could
be feedback, as \cite{kmt09b} suggest, or it could be the weight of the
{\rm HI} shielding layer, which has a regulatory effect on pressure
\citep{e89}. This regulatory effect works because at high ambient
pressure, the atomic density on the periphery of molecular clouds is
high and so the required surface column density for H$_{\rm 2}$ line
self-shielding is low, and vice versa. The pressure at the bottom of
the shielding layer, which is the molecular cloud surface pressure,
scales directly with the column density of the shielding layer. Thus a
lower intercloud pressure is compensated by a higher {\rm HI} column
density at the molecular cloud surface, making the molecular cloud
surface pressure somewhat uniform.

\cite{rk08} simulated star formation in galaxies. They took a star
formation rate per unit volume
\begin{equation}d\rho_{\rm stars}/dt = f_{\rm H2} \times (\rho_{\rm gas}/t_{\rm SF})
\times (n_{\rm H}/[10\; {\rm cm}^{-3}])^{0.5}\end{equation} where
$t_{\rm SF} = t_{\rm ff}/\epsilon_{\rm ff}$ is the free fall time,
$t_{\rm ff}$, divided by the fraction of the gas that turns into stars
in a free fall time, $\epsilon_{\rm ff}=0.02$. To determine the
molecular fraction, $f_{\rm H2}$, they considered heating and cooling,
a radiation field proportional to the SFR, the \cite{sternberg02}
H$_{\rm 2}$ formation theory, and radiative transfer using the {\it
Cloudy} code. The result was a SFR that scaled steeply with the total
gas column density, as observed, a higher KS slope for lower mass
galaxies, which is also observed, and a shallower KS slope for the
H$_{\rm 2}$ column density alone, as in the molecular KS law. These
results were somewhat independent of galaxy mass. The molecular/atomic
ratio also scaled with pressure in an approximately linear fashion,
regardless of galaxy mass, as observed. They also found a stability
parameter $Q$ that ranged from unstable in the inner, star-forming
parts of the disk, to stable in the outer regions.

\subsection{Observations and Models of Outer Disks}

\cite{murante10} have a multi-phase SPH code that assumes pressure
determines the molecular abundance, and the molecules give the SFR.
Below $\Sigma_{\rm total\;gas}\sim10\;M_\odot$ pc$^{-2}$, the slope of
the molecular star formation law turns out to be very steep,
$\Sigma_{\rm SFR}\propto\Sigma_{\rm total\;gas}^{n}$ for $n\sim4$.
Above $10\;M_\odot$ pc$^{-2}$, the slope is the same as in the
Kennicutt law, $n=1.4$, which is steeper than in \cite{bigiel08}, where
$n\sim1$ for the molecular Schmidt law.

\cite{bush10} simulated galactic star formation with special attention
to the outer disks. The star formation model followed \cite{sh02} with
radiative cooling, star formation in the cold phase, no specific
molecular phase, and a volume-Schmidt law, $\rho_{\rm
SFR}\propto\rho_{\rm total\;gas}^{1.5}$. They found patchy star
formation in the outer parts, usually along spiral arcs where the gas
density was high. This morphology is in agreement with GALEX
observations \citep{thilker05,gil05}. The Bigiel et al. and Leroy et
al. observations were matched qualitatively in these outer parts too:
below $\Sigma_{\rm gas}\sim10\;M_\odot$ pc$^{-2}$, the slope $n\sim6$
to 8 was steeper than in the observations (which also plot SFR versus
total gas in the outer regions). Then it was less steep at higher
column density, with a slope of $n\sim1.4$, which agrees with the
\cite{ken98} slope for total gas.

Outer disks can be Toomre-stable on average because the gas and star
column densities are very low. This is especially true for dwarf
galaxies \citep{hunter98}. It might be that magnetic fields and
viscosity destabilize outer disks, as discussed in Section
\ref{sect:q}, but in any case, outer disks appear to be much more
stable than inner disks. More importantly, the gas is outer disks is
often far from uniform and the use of an average column density for $Q$
is questionable. Locally there can be islands of high column density
where $Q$ is small enough to be in the unstable region \citep{vz96}.
These islands have to be larger than the Jeans mass, which might be
$10^7\;M_\odot$. Spiral arms and large disturbances in pressure could
make unstable regions like this.  \cite{dong08} found unstable islands
of star formation in the outer part of M83.  In these regions, the star
formation followed a steep KS law from point to point with a slope of
about 1.4 \citep{dong08}.

\cite{bigiel10a} found that outer disk star formation seen by GALEX
follows the {\rm HI} very well in M83, with a uniform consumption time
of $100$ Gyr per atom beyond $1.5R_{\rm 25}$. The form of the star
formation is mostly in spiral arms. Outer disk arms could be spiral
waves radiating from the inner disk \citep{ba10}.

\cite{boissier08} observed the galaxy-integrated KS law in low surface
brightness galaxies, using a SFR from GALEX NUV observations. For a
given total {\rm HI} mass, the star formation rate was low by a factor
of $\sim5$ compared to normal spirals, but over the whole range, the
total star formation rate scaled directly with the {\rm HI} mass. This
is not the same as saying that the star formation rate per unit area
scales directly with the {\rm HI} column density because the
observations are spread out in a plot like this with big galaxies on
one side and small galaxies on the other.

\cite{bigiel10b} studied SF rates in the far-outer disks of 17 spiral
and 5 dwarf galaxies, where the gas is highly {\rm HI} dominated. The
SF laws compare well with those in dwarf galaxies. There is no obvious
Q threshold. They suggest that the total SF Law has three components,
the extreme outer disk component that is {\rm HI} dominated, a
transition region where the molecular fraction increases to near unity,
the molecular region inside of that, and the starburst component, where
the surface density is higher than that of a single GMC.

\subsection{Scaling relations inside individual clouds}

\cite{kt07} showed that the conversion rate from gas to stars per unit
free fall time is about constant inside clouds over a wide range of
densities. This implies that the SFR per unit volume scales with the
1.5 power of density, with the first 1 in the power coming from the
mass per unit volume, and the 0.5 in the power coming from the free
fall rate. This is like a KS law, but for individual GMCs.  There is
also a threshold column density for star formation inside GMCs of
around $\sim5-7$ mag in V \citep{johnstone04,kirk06,enoch06, jorg07}.

\cite{chen10} studied the KS relation for individual GMCs in the LMC.
They measured the star formation rate from both {\rm HI}I regions and
by direct counting of young stellar objects. For YSO counting, the rate
per unit area inside a cloud approximately satisfies the total-gas
Kennicutt relation with the same time scale per atom, $\sim1$ Gyr. For
these regions, $\Sigma_{{\rm HI}+{\rm H}_{\rm 2}}\sim100\;M_\odot$
pc$^{-2}$, larger than in the main parts of galaxy disks. Chen et al.
also found that the areal rate of star formation was much lower in the
long molecular ridge south of 30 Doradus than in the GMCs.  Presumably
this ridge is not strongly self gravitating, even though it is {\rm
CO}-rich.

\section{Summary}

The empirical star formation law on kpc scales is essentially one where
star formation follows {\rm CO}-emitting molecular gas with a constant
rate per molecule, and the ratio of molecular to atomic gas scales
nearly directly with the ISM pressure \citep{bigiel08,leroy08}. The
rate per molecule corresponds to a consumption time of molecular gas
equal to about 2 Gyr. The place in a galaxy where the transition occurs
between {\rm HI} dominance in the outer part to H$_{\rm 2}$ dominance
in the inner part is at a pressure of $P=2.3\pm1.5\times10^4k_{\rm B}\;
{\rm K}\; {\rm cm}^{-3}$. There also tend to be characteristic gas and
stellar column densities at this place, and a characteristic galactic
orbit time for all of the galaxies observed. Beyond this radius is the
atomic-dominated outer disk. There, the SFR scales directly with
$\Sigma_{\rm {\rm HI}}$, and the consumption time is about 100 Gyr.

Theoretical models of these empirical laws include the
atomic-to-molecular transition in individual clouds and a sum over
clouds to give the galactic scaling laws.  Star formation occurs only
in the densest parts of the clouds, as determined by a combination of
turbulence-compression and self-gravity. Numerous simulations of star
formation in galaxies can reproduce these empirical laws fairly well.
The simulations usually show a sensitivity to the Toomre $Q$ parameter,
unlike the observations.


\end{document}